\documentclass[11pt,preprint,preprintnumbers,amsmath,amssymb,nofootinbib]{revtex4}
\usepackage{color}
\usepackage{graphicx,graphics}

\newcommand{\be}{\begin{equation}}  
\newcommand{\ee}{\end{equation}}  
\newcommand{\bea}{\begin{eqnarray}}  
\newcommand{\eea}{\end{eqnarray}}  
\newcommand{\ba}{\begin{array}}  
\newcommand{\ea}{\end{array}}

\begin{document}
\preprint{FERMILAB-PUB-21-074-T}
\preprint{OUTP-21-067P}

\title {Starobinksy Inflation, Gravitational Contact Terms 
\\and the  Induced BEH Boson Mass}


\author{Christopher T. Hill}
\email{hill@fnal.gov}
\affiliation{Fermi National Accelerator Laboratory\\
P.O. Box 500, Batavia, Illinois 60510, USA\\$ $}
\author{Graham G. Ross}
\email{g.ross1@physics.ox.ac.uk}
\affiliation{Rudolf Peierls Centre for Theoretical Physics, \\
University of Oxford, 1 Keble Road\\
Oxford OX1 3NP\\$ $}

\date{\today}
\begin{abstract}
The Starobinsky model of inflation remains consistent with observation, forty years after its  introduction. It provides a well motivated origin for the scalar inflaton, the ``scaleron'' with a mass of $O(10^{13})$ GeV, emerging as a graviton degree of freedom from $R^2$ corrections to Einstein-Hilbert gravity.  However the coupling of such a heavy state to the  BEH (``Higgs'') scalar is problematic as its quantum loop corrections can induce an unacceptably large contribution to the radiatively induced BEH  scalar mass.  The calculation of these corrections is normally done by Weyl transforming to the Einstein frame, yet at the quantum level Weyl transformations are fraught with ambiguities.  However the recent realization that there exist ``gravitational contact interactions'' largely sidesteps these ambiguities. Such contact terms are necessarily present, coming from t-channel graviton exchange interactions, and they show that the theory defined in the ``Jordan Frame" is identical to the theory in the Einstein frame, with additional Planck-scale suppressed interactions that take on the form of a Weyl transformation. 
This avoids ambiguous nonlinear field redefinitions, and reliable loop calculations are possible leading to a consistent low energy theory in an expansion in $1/M^2$. Taking account of the contact terms 
we study the radiative corrections to the BEH mass in the mixed Higgs/$R^2$ model with explicit scale breaking,  and in an extension of the model in which exact scale symmetry is spontaneously broken.
\end{abstract}

\maketitle

\section{Introduction}\label{intro}
There has been considerable interest in recent years in the
fundamental role of scale symmetry with gravity and the cosmological
evolution of the early universe.
Scale, or equivalently Weyl invariance (the natural setting for scale symmetry when gravity is incorporated), offers a class
of models in which cosmic inflation, the origin of
the Planck mass as a dynamical scale symmetry breaking, 
and potentially large hierarchies seen in nature, can
arise as a unified phenomenon.

In the present paper we focus on a
leading candidate quantum field theory of inflation known as  the
Starobinsky Model \cite{Starobinsky:1980te} both in its original form with explicit scale breaking,  and in an extension of the model in which exact scale symmetry is spontaneously broken.  This is an $R^2$ theory, and as such has an
extra physical scalar graviton degree of freedom beyond the 
two propagating degrees of freedom in the Einstein-Hilbert theory.
One can immediately ``factorize'' the $R^2$ interaction, replacing it with a scalar auxiliary field Lagrangian, $\sim \eta^2 R - \xi \eta^4$, and $\eta$ will 
subsequently acquire kinetic terms.
This new field, $\eta$, is
known as the ``scaleron'',  emerges automatically
in the model and  can serve as the inflaton.  This in turn
leads to a phenomenologically successful inflationary model provided the scaleron is sufficiently heavy of $O(10^{13})$ GeV.
We are interested in the effective quantum field theory aspects of this scheme. In particular we explore the naturalness
of the light Brout--Englert--Higgs (BEH, ``Higgs'') scalar \cite{Englert:1964et,Higgs:1964ia,Higgs:1964pj} of the Standard Model when it is
introduced into the theory and interacts with the scaleron state, the ``Higgs/$R^2$" model. This
is particularly interesting in the scale invariant version of the model because scale invariance forbids a bare BEH mass and it is only generated on spontaneous breaking of the symmetry.


A standard technology used in performing loop calculations in effective
field theory in these schemes involves the ``Weyl transformation.''
While an exact transformation classically, the Weyl transformation involves
nonlinear field redefinitions, and leads to a morass of ambiguities at the quantum level.
%
Mainly, Weyl transformations can transform a  classical theory
with nonminimal couplings of scalars to gravity, as $F(\phi)R$,
into any other coupling, $G(\phi)R$.  There is a special frame, $G(\phi)
= M^2$, called the ``Einstein frame,'' where the nonminimal couplings are absent.  However,  one would think any frame should be as good as any other
and indeed some authors invoke special properties of
other frames, such as those with ``conformal coupling, ''
$-\frac{1}{12}\phi^2 R$.  

Recently, however,  we pointed out that if one considers the $t$-channel exchange of gravitons in a  theory with Planck scale and non-minimal interactions, $(M^2 + F(\phi))R$, a novel  phenomenon occurs \cite{Hill:2020oaj}: the theory has contact term interactions. In weak field approximation, $g_{\mu\nu} \approx\eta_{\mu\nu} + h_{\mu\nu}/M$ the scalar curvature has a leading $F(\phi) \partial^2 h $
term.\footnote{This term is projected to zero in the Einstein-Hilbert
action where $ \int d^4x\sqrt{-g} M^2 D^2h$ is a total divergence}
In a Feynman diagram this implies a vertex proportional to $q^2$,
which in turn cancels the $1/q^2$ in the  graviton propagator. This means that the would-be long range graviton exchange potential arising from the non--minimal terms  collapses to a delta-function,
and becomes a set of local operators.   The contact term 
integrates out the 
$ F(\phi)R$ term and replaces it with higher dimension operators, $\sim F D^2 F/M^2$ and $\sim F T_\mu^\mu/M$.

In exploring radiative corrections we find, if we do not take the contact term into account, the resulting induced
BEH scalar mass is dependent upon the frame choice.  Effectively the different Weyl frames, as conventionally treated in the absence of the contact term, are then {\em different theories}. 
Therefore, it would appear  that Weyl transformations 
are incompatible with quantum theory.  Indeed, this incompatibility has been
argued to be the case long ago by M. Duff  \cite{Duff:1980ix}.
This issue has been addressed by \cite{Steinwachs},
who find that the frame dependence is remedied by working on-mass
shell for background fields, i.e., by imposing equations of motion.
 But what of off-shell quantities, such as $\beta$-functions?  How is
 uniqueness of the theory realized as an action?

 It is important to realize that  the
contact term, in the $1/M_P^2$ expansion, induces
the higher dimension local operators {\em that
are part of the effective action.}   In effect, the contact term
induced operators
are hidden parts of the full effective action in any given frame.
In any putative  ``Jordan frame'' the non-minimal interaction will always be replaced  by these new operators.
The contact term in gravity, though it involves
no field redefinitions, has the same structure in any frame as a Weyl transformation leading to the Einstein frame\footnote{In reference \cite{Hill:2020oaj} the leading behaviour
in $1/M^2$ was derived, but presumably the t-channel diagrams
can be resummed to yield the full Weyl transformation form. While we have not explicitly proved this, one can see evidence of it by considering multi-graviton exchange diagrams, or considering iteration of the Einstein
equation for $R$.}. 
However, it is important to note that  the contact term is not a Weyl transformation. 

Due to the contact terms we see that the Jordan frame, which has anomalous couplings of the scalars to the Ricci scalar,  is  identical to the Einstein frame with only an Einstein Hilbert term, $M_P^2R$, and higher dimension operators.  The contact terms can be seen as arising by 
judiciously using
the equation of motion for $R$ within the non-minimal terms, 
hence it is
not surprising that the consistency amongst frames observed
in \cite{Steinwachs} occurs on-shell. There the imposition of
the equantion of motion implicitly implements the contact terms.
However, the contact term assures that the $\beta$-functions computed
in the Einstein frame are the unique and correct ones.

 Contact terms are familiar
elsewhere in physics. A well-known example occurs in electroweak
phsyics, where we have ``penguin diagram,'' in  which electroweak
vertex corrections to gluon exchange generates contact terms.
These 
lead to new local 4-fermion operators
which mediate processes such as $K\rightarrow 2 \pi$.
The contact term is
classical, of ${\cal{O}}(\hbar^0)$ arising from tree diagrams, yet contact terms are an exception to the rule
that corrections to the action come only from single particle irreducible (loop) diagrams.  The contact term is an ab initio
hidden yet essential part of the physics.  Moreover, one would be 
double counting to add the contact terms to the action without
removing the $q^2$ vertices as well. 

One still has an ambiguity of choice of source currents in
an effective potential calculation in the Einstein frame an example is given in Appendix C.  This is a ``user option'',
analogous to whether one computes a magnet's potential as function 
$\overrightarrow{M}$ (magnetization) or $\overrightarrow{M}^2$.  Many choices are possible, and lead to physically different potentials and interpretations.
Presently we are interested in the order parameter that
can serve optimally as the inflaton.  
In studying inflation it is necessary to consider the evolution of fields away from equilibrium and motivates  one 
to choose a variable that diagonalises the kinetic term in the absence of a constant background field.  For the original Starobinsky model this leads one to a natural fundamental variable eliminating the remaining ambiguity.

In calculating the effective action in the Coleman-Weinberg
method \cite{Coleman:1973jx}, it is important to realize
that the source terms lead to a 
result with  ``on-shell'' classical background fields.
Merely shifting and computing without sources
can lead to results that are dependent on the choice 
quantum fluctuations (the actual field we integrate in the path integral). For potentials, ``on-shell'' means that static classical background fields are VEVs.
 Since we are interested in the effective potential far
 from the true potential minimum we add the source terms that shift these
 fields to arbitrary classical values. These shifted fields are then minima of  the {\em full potential with sources}, where the currents have deformed  the potential. 
 In the special case that background fields {\it are} in the true local
 minimum of the classical potential, where the sources vanish,
 then the fields are  automatically on-shell.
 With a given set of sources and  on-shell
 background fields, the effective potential will then be invariant under different choices of the quantum fluctuation fields. A proof and discussion of this is given in Appendix  \ref{onshell}.

{ We also introduce an alternative method to compute the effective potential using the renormalization group (RG). This is an expeditious method based on the fact that the rather complicated 
and nonrenormalizable form of the potential at short-distance
 (Planck-scale)
 contains a subset of ``relevant operators'' that propagate logarithmically into the infrared and define the low energy theory. We write down a generic set of relevant operators, match
 their coefficients at $M$ to the Starobinsky potential, and compute the evolution
 of their couplings into the infrared with conventional RG equations.  This
 is similar to a second order phase transition in condensed matter physics, where there is approximate scale invariance
 and sensitivity to the complicated short-distance physics is erased,
 leading to universal low energy results.
 Presently, the low energy form of the Higgs mass is 
 found to be identical to
 the Coleman-Weinberg form and can be
 directly connected to the trace anomaly \cite{cth}.  This is a powerful and efficient technique, and will
 be developed further elsewhere \cite{HRelsewhere}.}

Contact terms thus mandate a unique and well defined formulation of the theory in the Einstein frame.
In what follows we will apply this to focus upon two settings for the Starobinsky model. 
We will begin with a non-scale-invariant theory in which the Planck mass is present as a fundamental scale and the model is extended to include the BEH scalar. 
We will  also analyse a Weyl invariant version of the Starobinsky model including the BEH scalar
in which the fundamental mass scales arise by spontaneous symmetry breaking \cite{Brans:1961sx}. This requires an additional scalar field and forms the basis of the modern Weyl invariant field
theory approach  \cite{ShapoZen,ShapoZen2,ShapoBlas,GarciaBellido:2011de,Ferreira:2016vsc,Ferreira:2016wem,Salvio:2017qkx,
Wang:2017fuy,Ema:2017rqn,He:2018gyf,Gundhi:2018wyz,Enckell:2018uic,He:2020ivk,Ghilencea:2018rqg,Ghilencea:2018dqd}.

Here the spontaneous breaking of the Weyl-scale symmetry is 
 a consequence of the conserved  Weyl current, $K_\mu =\partial_\mu K$
where $K$ is a scalar function of the scalar fields.  Any conserved
 current  will redshift to zero in a general expansion of the universe,
 and so too the Weyl current.
 However, since the Weyl current is a derivative of a
 scalar $K$, we see that $K$ will therefore redshift to an
 arbitrary constant, $\bar{K}$.  
 This is the order parameter of the spontaneous symmetry breaking,
 and when it acquires a nonzero VEV the symmetry is broken.
 We call this ``inertial symmetry breaking'' \cite{Ferreira:2018itt} because it does  not involve a potential.
 There will be a dilaton, and the decay constant of the
 dilaton is $\sqrt{2\bar{K}}$, and is also proportional to
 the Planck mass.

\section{The  mixed Higgs/$R^2$ Model}\label{NSI}

\subsection{The Action}\label{NSI1}

We focus presently 
upon the mixed Higgs/$R^2$ Model that extends the
Starobinsky inflationary model  \cite{Starobinsky:1980te} to include the BEH boson.
The model has the action:
\bea
\label{R2S}
S &=& \int\sqrt{-g}\left(\frac{1}{2}g^{\mu\upsilon}\partial_{\mu}{H} \partial_{\nu
}{H}-\frac{1}{6f_{0}^{2}}R^{2}-\frac{1}{12}\alpha_H 
H^2 R+\frac{1}{2}%
M^{2}R-V(H) \right)
\eea
where $M$ is the Planck mass and ${R}({g})$ is the Ricci scalar of the metric $g_{\alpha\beta}$.  
Here we also include the Standard Model BEH scalar isodoublet 
$\mathcal{H}$, which, for convenience
we will treat $\mathcal{H}$, as a real scalar field $H$. This
can be identified with the physical BEH isodoublet scalar of the Standard Model 
when  $\mathcal{H}$ is written in the 
unitary gauge $\mathcal{H}=(0,H/\sqrt{2})$ and we assume
the the electroweak gauge fields are pure-gauge configurations. Then
$D_\mu\mathcal{H}\rightarrow \partial_\mu H$ and $\mathcal{H}^\dagger\mathcal{H}
\rightarrow\frac{1}{2} H^2$.
We have allowed for  a non-minimal coupling of $ H$ proportional to $\alpha_H$. Even if this is initially set to zero it is generated radiatively by Standard Model couplings \cite{Herranen:2014cua,Herranen:2015ima}.  
We will largely ignore $V(H)$ as it plays a subdominant role in the radiative corrections.

Since the $R^2$ term involves fourth order derivatives it contains an additional (scalar) degree of freedom \cite{Whitt:1984pd,Hindawi:1995an}. To make this explicit it is conventional to reduce the fourth order derivatives to second order by introducing a static
auxiliary field, $\eta$,  with the action now given by:
\bea
S=\int\sqrt{-g}\left(\frac{1}{2}g^{\mu\upsilon}\partial_{\mu}%
H\partial_{\nu}H-\frac{1}{12}\alpha_{\eta}\eta^{2}R-\frac{1}{12}\alpha
_{H}H^{2}R+\frac{1}{2}M^{2}R-\frac{\xi^{\prime}}{4}\eta^{4}
\right)
\label{action1}
\eea
where the equation of motion yields: 
\bea
\eta^{2}= -\frac{1}{6}\frac{\alpha_\eta}{\xi^\prime} R 
\qquad \qquad \makebox{and:} \qquad \qquad
 f_{0}^{2}= \frac{24\xi^\prime }{\alpha_{\eta}^{2}}.
\eea
The coupling $\xi^{\prime}$ and the value of $\alpha_{\eta}$ are relative.
We can define a new $\xi=\xi^{\prime}/\alpha_{\eta}^{2}$  and rescale
$\eta^{2}$ so that $\alpha_{\eta}=1$ \ and,
\bea
-\frac{1}{12}\alpha_{\eta}\eta^{2}R-\frac{\xi^{\prime}}{4}\eta^{4} \;
\longrightarrow \; -\frac{1}{12}\eta^{2}R-\frac{\xi}{4}\eta^{4}.
\eea
We will take $\alpha_{\eta}^{2}=1$
to be ``standard normalization'', in which case we have 
$f_{0}^{2}= {24\xi }$.
In the standard normalization the action becomes:
\bea
S=\int\sqrt{-g}\left(\frac{1}{2}g^{\mu\upsilon}\partial_{\mu}%
H\partial_{\nu}H+\frac{1}{2}M^{2}R\Omega^{2}\ \ -\frac{\xi}{4}\eta^{4}%
\right)
\eea
where we define:
\bea
\label{sixx}
\Omega^{2}=  1-\frac{\alpha_{H}}{6M^{2}}%
H^{2}-\frac{1}{6M^{2}}\eta^{2}  \equiv  \exp\left(  \frac{2\chi
}{\sqrt{6}M}\right)  
\eea
and $\chi$ will play the role of the inflaton.

While this  formulation is done in the ``Jordan frame,'' as shown in \cite{Hill:2020oaj},  contact interactions generated by tree level graviton exchange significantly modify the action.  In this sense the Jordan frame does not really exist, as it is driven to the Einstein frame by the contact terms. In practise, rather than carrying out the laborious process of calculating the contact terms, we can formally perform a Weyl
transformation to go to an Einstein-Hilbert action: 
\bea
g_{\mu\nu}(x)&\rightarrow & \Omega^{-2}g_{\mu\nu}(x) \qquad 
g^{\mu\nu}(x)\rightarrow\Omega^{2}g^{\mu\nu}(x)
\nonumber \\
\sqrt{-g}&\rightarrow&\sqrt{-g}\;\Omega^{-4}
\qquad
R(\Omega^{-2}g)=\Omega^{2}R(g)+6\Omega^{3}D\partial\Omega^{-1}.
\eea
We emphasize however that {\em the contact terms are not a Weyl transformation}
and preserve the original
metric, though formally the resulting structure of
the theory is that of the Weyl transformation.

Then we have from eq.(\ref{sixx}),
\bea
\eta^{2}=6M^{2}\left(  \exp\left(  \frac{2\chi}{\sqrt{6}M}\right)  -\left(
1-\frac{\alpha_{H}}{6M^{2}}H^{2}\right)  \right). 
\eea
and the potential takes the form:
\bea
\label{pot17}
\frac{\xi}{4}\Omega^{-4}\eta^{4}&=&\frac{3}{8}M^{4}f_{0}^{2}
\left(
1-\exp\left( -\frac{2  \chi  }{\sqrt{6}M}\right)
\left(  1-\frac{\alpha_{H}}{6M^{2}}H^{2}\right)  \right)  ^{2}.
\eea

We now turn to two different treatments of the effective potential.
One follows the Coleman-Weinberg approach, introducing sources
and performing an expansion to order $\hbar$.
The second approach uses the renormalization group
and is much simpler.

\subsection{A Coleman-Weinberg Calculation of the Effective Potential}

The Coleman-Weinberg
potential \cite{Coleman:1973jx} results from a WKB approximation to computing
a Gaussian integral in field theory.  We
add source terms to dynamically shift the classical values of the fields 
to nonzero on-shell VEVs. Presently, we choose a source term
$ J\sqrt{6}M\ln\Omega+KH $. The
parameterization of the fields is then arbitrary and we will assume,
$
\chi=\sqrt{6}M\ln\Omega $ 
and
$H$, and then note that
$ 3M^{2}\partial
\ln\Omega\partial\ln\Omega=\allowbreak\frac{1}{2}\partial_{\rho}\chi
\partial^{\rho}\chi.$
The Einstein frame action is then:
\bea
\label{nine}
S&=& \int\sqrt{-g}\left(\frac{1}{2}\Omega^{-2}\partial_{\rho}H\partial^{\rho}%
H+\frac{1}{2}\partial_{\rho}\chi\partial^{\rho}\chi+
\frac{1}{2}M^{2}R
-\frac{\xi}{4}\Omega^{-4}\eta^{4}
+J\sqrt{6}M\ln\Omega+KH\right)
\eea
We emphasize that eq.(\ref{nine}) 
is a result of the contact terms, and the metric
appearing is the original metric of eq.(\ref{action1}).

We now expand in in $\sqrt{\hbar}\equiv \epsilon$:
\bea
H=h+\epsilon x,\qquad\chi=\kappa+\epsilon y, \qquad \sqrt{\hbar}=\epsilon
\eea
where $(h,\kappa)$ are classical background
fields and $(x,y)$ are quantum fluctuation
scalar fields, integrated in the  the path integral.
We also introduce the following useful
functions of the classical background fields:
\bea
P=\exp\left(-\frac{\kappa}{\sqrt{6}M}\right),  \qquad\qquad p=\sqrt
{\frac{\alpha_{H}}{6M^{2}}}h, \qquad\qquad q=\sqrt{\frac{1}{6M^{2}}}h.
\eea
The action to \ O$\left(  \hbar\right)  =\epsilon^{2}$ \ becomes:
\bea
S=\int\sqrt{-g}\left(\frac{1}{2}M^{2}R+\frac{1}{2}\epsilon^{2}%
P^{2}\partial_{\rho}x\partial^{\rho}x\ +\frac{1}{2}\epsilon^{2}\partial_{\rho
}y\partial^{\rho}y-V(J,K)\right),
\eea
where the potential of eq.(\ref{pot17}) including sources is given by: 
\bea
V(J,K)=
\frac{\xi}{4}\Omega^{-4}\eta^{4} +J\left(  \kappa+\epsilon y\right)  +K\left(
h+\epsilon x\right),
\eea
where,
\bea
& & \frac{\xi}{4}\Omega^{-4}\eta^{4}=
\nonumber \\
&& 
\!\!\!\!\!\!\!
\frac{3}{8}M^{4}f_{0}^{2}\left(  1-P^{2}\left(  1-\frac{2}{\sqrt{6}%
M}y\epsilon+\frac{1}{3M^{2}}y^{2}\epsilon^{2}+O\left(
\epsilon^{3}\right)  \right)  \left(  1-\frac{\alpha_{H}}{6M^{2}}\left(
\allowbreak h^{2}+2h\epsilon x+\epsilon^{2}x^{2}\right)  \right)  \right)
^{2}.
\eea
The sources are determined by the condition that the potential
$V(J,K)$ be evaluated at the minimum, whence the linear terms in $\epsilon$ vanish.
Since the particular sources we have chosen
couple linearly to the fields they do not affect
the quadratic term, $O\left(  \epsilon^{2}\right)  $,  in the expansion of
the  potential (see Appendix B for an exception). 
Then performing the Legendre transformation:  
\bea
V(\kappa,h)=V(J,K)-J\frac{d}{dJ}V(J,K)-K\frac{d}{dJ}V(J,K)
\eea
with  $G=\frac{3}{8}M^{4}f_{0}^{2}=\allowbreak9M^{4}\xi$,
we expand $V(\kappa,h)$ to  $O\left(  \epsilon^2\right)$ 
to obtain the potential, $V(\kappa,h)$,
\bea
V(\kappa,h) &=& V_{classical}
\nonumber \\
&& +\frac{G}{M^{2}}\epsilon^{2}\left(
\frac{1}{3}\left(  1-P^{2}\left(  1-p^{2}\right)  \right)  P^{2}\left(
\alpha_{H}x^{2}-4\sqrt{\alpha_{H}}pxy-2y^{2}\left(  1-p^{2}\right)  \right)\right.
\nonumber \\
&&
+\frac{2}{3}P^{4}\left(  \sqrt{\alpha_{H}}px+y\left(  1-p^{2}\right)
\right)^{2}%
\eea
where the classical potential is 
$V_{classical}=
(  1-P^{2}\left(  1-p^{2}\right) )  ^{2}$.

\subsection{One loop radiative corrections}
At $O(\hbar)$, the quadratic term in the fluctuations can be written as an unrenormalized mass squared matrix 
amongst the quantum fluctuations $\ \left(  x,y\right)  $:
\bea
&&
\frac{1}{2}\left(
\begin{array}
[c]{c}%
y\\
x
\end{array}
\right)^{T}\Xi_{0}^{2}\left(  \
\begin{array}
[c]{c}%
y\\
x
\end{array}
\right)\qquad  \makebox{where},
\nonumber \\
&& \Xi_{0}^{2}=\frac{2G}{M^{2}}\left(
\begin{array}
[c]{cc}%
\frac{2}{3}P^{2}\left(  p^{2}-1\right)  \left(  1-2P^{2}\left(
1-p^{2}\right)  \right)   & \;\; -\frac{2}{3}P^{2}%
\sqrt{\alpha_{H}}p\left(  1-2P^{2}\left(  1-p^{2}\right)  \right)  \\
\allowbreak-\frac{2}{3}P^{2}\sqrt{\alpha_{H}}p\left(  1-2P^{2}\left(
1-p^{2}\right)  \right)   & \;\; \frac{1}{3}P^{2}\alpha
_{H}\left(  1-P^{2}+3P^{2}p^{2}\right)
\end{array}
\right)  .
\eea
Likewise the kinetic terms for $\left(  x,y\right)  $ can be written:
\bea
\frac{1}{2}\epsilon^{2}P^{2}\partial_{\rho}x\partial^{\rho}x\ +\frac{1}%
{2}\epsilon^{2}\partial_{\rho}y\partial^{\rho}y=\left(
\begin{array}
[c]{c}%
\partial y\\
\partial x
\end{array}
\right)  ^{T}  Z \left(
\begin{array}
[c]{c}%
\partial y\\
\partial x
\end{array}
\right)  
\eea
where,
\bea
Z=\left(
\begin{array}
[c]{cc}%
\allowbreak1\allowbreak & 0\\
0 & \allowbreak P^{2}%
\end{array}
\right)  \qquad\qquad\sqrt{Z}^{-1}=\left(
\begin{array}
[c]{cc}%
\allowbreak1\allowbreak & 0\\
0 & P^{-1}\allowbreak\allowbreak
\end{array}
\right) .
\eea
The physical, renormalized mass squared matrix is therefore:
\bea
\Xi^{2}&=&\sqrt{Z}^{-1}\Xi_{0}^{2}\sqrt{Z}^{-1}
\nonumber \\
&=&\frac{8}{3}\frac{G}{M^{2}}\left(
\begin{array}
[c]{cc}%
-\frac{1}{2}P^{2}\left(  1-p^{2}\right)  \left(  1-2P^{2}(1-p^{2}\right)   &\;\;
-\sqrt{\alpha_{H}}pP\left(  \allowbreak\frac{1}{2}-P^{2}(1-p^{2})\right)  \\
-\sqrt{\alpha_{H}}pP\left(  \allowbreak\frac{1}{2}-P^{2}(1-p^{2})\right)   &\;\;
\frac{1}{4}\alpha_{H}\left(  1-P^{2}(1-3p^{2}\right)  ).
\end{array}
\right)  
\eea
Strictly following the analysis of Coleman and Weinberg \cite{Coleman:1973jx}  we should construct the eigenvalues of the mass squared matrix, 
denoted as $( \Xi_1^{2}, \Xi_2^{2})$. The potential then takes the form, subject to renormalization conditions,
\bea
\Gamma = -\frac{1}{64\pi^2}\Xi_1^{4}\ln(M^2/\Xi_1^{2}) -\frac{1}{64\pi^2}\Xi_2^{4}\ln(M^2/\Xi_2^{2}),
\label{CW}
\eea
(see Appendix A).
Here the  mass eigenstates correspond to the scaleron and the BEH scalar. Given the large mass of the scaleron it dominates the contribution to eq(\ref{CW}) and to an excellent approximation we can can define
the resulting effective potential as:
\bea
\Gamma=-\frac{1}{64\pi^{2}}\makebox{Tr}\left(  \Xi^{4}\right)  \left(  \ln\left(
M^{2}/m^{2}\right)  \right)
\label{EP}  
\eea
where $m$ is the scaleron mass 
 and
we have:
\bea
\makebox{Tr}\left(  \Xi^{4}\right) 
&=&\frac{16}{9}\frac{G^{2}}{M^{4}}\left(
P^{4}\left(  1-p^{2}\right)  ^{2}\left(  1-2P^{2}\left(  1-p^{2}\right)
\right)  ^{2}+8\allowbreak\alpha_{H}p^{2}P^{2}\left(  \frac{1}{2}-P^{2}\left(
1-p^{2}\right)  \right)  ^{2}\right.
\nonumber \\
&&\qquad\qquad \left.
+\frac{1}{4}\alpha_{H}^{2}\left(  1-P^{2}\left(  1-3p^{2}\right)  \right)
^{2}%
\right) 
\eea
This is the leading order result in $\hbar=\epsilon^{2}$ valid to
all orders in $(\kappa,h)$.
We are interested in calculating the correction to the BEH mass at the end of inflation in the limit of small  $\kappa\rightarrow0$, hence $P=\exp\left(-\frac{\kappa}{\sqrt{6}}\right)  \rightarrow1$  and
\bea
\makebox{Tr} \left(  \Xi^{4}\right)  \rightarrow
\frac{16}{9}\frac{G^{2}}{M^{4}}\left(  1+\left(
-\frac{\alpha_{H}}{M^{2}}+\frac{1}{3}\frac{\alpha_{H}^{2}}{M^{2}}\right)
h^{2}+O\left(  h^{4}\right)  \right)  
\eea
and,
with ${\ f_{0}^{2}}={24\xi}$, 
$G=({3}/{8})M^{4}f_{0}^{2}=9M^{4}\xi$, the loop induced Higgs mass term is then:
\bea
\Gamma=-\frac{3}{4\pi^{2}}M^{2}\xi^{2}\alpha_{H}(\alpha_{H}-3)h^{2}\ln\left(
M^{2}/m^{2}\right)  =-\frac{1}{768\pi^{2}}M^{2}f_{0}^{4}\alpha_{H}\left(
\alpha_{H}-3\right)  h^{2}\left(  \ln\left( M^{2}/m^{2}\right)
\right)  
\label{mt1}
\eea
Note the parameters refer to the action of eq(1) where $\xi$ is rescaled 
to standard normalization.  This is our key result for the NSI Starobinsky model, and  disagrees with previous results, such as \cite{Salvio:2017qkx}.

\subsection{Renormalization Group Approach}

The effective potential approach used in the preceding section is a
well-defined procedure. Its relationship to other fundamental
quantities in the field theory, such as the $\beta$-functions
and trace anomalies, is not so obvious.  However, we  can easily obtain the preceding results
with the renormalization group equations. The connection to
the trace anomaly is then direct, as in ref.\cite{cth}.

We construct a  generic
theory consisting of ``relevant operators''
up to dimension 4, that participate in the logarithmic running
of the relevant coupling constants.
We  then impose boundary conditions
at the UV (Planck) scale that match the  parameters
of this generic potential
to the non-scale invariant Starobinsky model.

Consider the following generic field theory:
\bea
S=\int \frac{1}{2}\partial_{\rho}H\partial^{\rho}H+\frac{1}{2}\partial_{\rho}\chi\partial^{\rho}\chi-W(\chi,H)
\eea
where the potential is
\bea
W(\chi,H)=\frac{1}{2}\mu_{1}^{2}\chi^{2}+\frac{1}{2}\mu_{2}^{2}H^{2}+m_{1}%
\chi^{3}+m_{2}\chi H^{2}+\frac{\lambda_1}{4}\chi^{4}+\frac{\lambda_2}%
{4}H^{4}+\frac{\lambda_3}{2}H^{2}\chi^{2}.
\eea
This maintains the symmetry $H\rightarrow -H$.
We expand the fields $(\chi,H)$ in VEVs + fluctuations,
$(\chi,H)\rightarrow (\kappa+\hbar x,h+\hbar y)$
and integrate the fields $(x,y)$ to obtain one-loop
RG equations (this can be done efficiently
by the $\makebox{Tr}(\Xi^4)$
method 
as in Appendix A).
There are then seven renormalization group
equations in terms of
$(t = \ln(\mu))$:\footnote{We specialize to the case $P=1$
which would otherwise lead to wave-function renormalization corrections.
We also presently ignore the interesting issue of $\alpha_i$ running
which necessitates re-generation of the contact term.
These issues will be dealt with in a subsequent paper \cite{HRelsewhere}.}
\bea
\label{RGeqs}
8\pi^2\frac{\partial}{\partial t}\mu_{2}^{2}& =& 
-\left(3\lambda_2\mu_2^2+4m_{2}^{2}+\lambda_3\mu_{1}^{2}\right)
\nonumber \\
8\pi^2\frac{\partial}{\partial t}
\mu_{1}^{2}&=& -\left( 2 m_{2}^{2}+\lambda_3%
\mu^{2}+18m_{1}^{2}+3\lambda_1\mu_{1}^{2}\right)
\nonumber \\
8\pi^2\frac{\partial}{\partial t}
m_{1}&=&- \left(9\lambda_1m_{1}+  \lambda_3m_{2}%
\right)  
\nonumber \\
8\pi^2\frac{\partial}{\partial t}
m_{2}&=& \left(  3\lambda_2m_{2}+4\lambda_3%
m_{2}+3\lambda_3m_{1}\right)  
\nonumber \\
8\pi^2\frac{\partial}{\partial t}
\lambda_1&=&\left(  
9\lambda_1^{2}+\lambda_3^{2} \right)  
\nonumber \\
8\pi^2\frac{\partial}{\partial t}
\lambda_2&=&\left(  9\lambda_2^{2}%
+\lambda_3^{2}\right)  
\nonumber \\
8\pi^2\frac{\partial}{\partial t}
\lambda_3&=&
\lambda_3\left(  3\lambda
_{2}+4\lambda_3+3\lambda_1\right).  \qquad\qquad\qquad
\eea
While these can be formally integrated in 
leading log approximation, here we are only interested in
the single log solutions, such as,
\bea
\lambda_1 = \lambda_{1_0}-
\frac{1}{8\pi^2 }
\left(  \lambda_{3_0}^{2}
+9\lambda_{1_0}^{2}\right) \ln\left(\frac{M}{\mu}\right)
\eea
where $\lambda_{i_0}$ is the initial value of $\lambda_{i}$
at $\mu = M$ (see Appendix A).

To obtain the Higgs mass we use the NSI potential
as the boundary condition on the theory when $\mu=M$.
From eq(\ref{pot17}):
\bea
\frac{\xi}{4}\Omega^{-4}\eta^{4}&=&\frac{3}{8}M^{4}f_{0}^{2}
\left(
1-\exp\left( -\frac{2\kappa   }{\sqrt{6}M}\right)
\left(  1-\frac{\alpha_{H}}{6M^{2}}h^{2}\right)  \right)  ^{2}
\eea
Expanding this to quartic order in the fields and matching coefficients,
our generic potential takes the form at the Planck-scale:
\bea
W(\kappa,h)/ f_0^2=\frac{1}{4}M^{2}\kappa^{2}-\frac{1}{2\sqrt{6}}M\kappa^{3}%
+\frac{\alpha_{H}}{4\sqrt{6}}M\kappa h^{2}-\frac{\alpha_{H}}{8}\kappa
^{2}\allowbreak h^{2}+\frac{7}{72}\kappa^{4}+\frac{1}{96}\alpha_{H}^{2}h^{4}\eea 
In particular we see that, at the Planck-scale, $M$,
the initial values of the parameters relevant to the Higgs mass are:
\bea
\mu_{2_0}^2 = 0\qquad m_{2_0}=\frac{\alpha_{H_0}}{4\sqrt{6}}Mf_0^2
\qquad \mu_{1_0}^2 = \frac{1}{2}M^2f_0^2\qquad \lambda_{3_0} = -\frac{1}{4}\alpha_{H_0}f_0^2
\eea
Therefore, combining these results with the first line of eqs.(\ref{RGeqs}), the Higgs mass
is determined at a lower energy scale $\mu=m$ in the single log approximation:
\bea
\mu_{2}^{2}(\mu)
& =&-\frac{1}{384\pi^{2}%
}f_{0}^{4}M^{2}\alpha_{H}\left(  \alpha_{H}-3\right)  \ln\left(M^2/m^2\right)
\eea
yielding the Higgs potential:
\bea
\Gamma = \frac{1}{2}\mu_{2}^{2}h^2=
-\frac{1}{768\pi^{2}%
}f_{0}^{4}M^{2}\alpha_{H}\left(  \alpha_{H}-3\right)h^2  \ln\left(M^2/m^2\right)
\eea
in exact agreement with eq(\ref{mt1}).

We have presented the renormalization group (RG) approach here to
illustrate that it is much
simpler and potentially more
powerful than the effective potential calculation obtained
previously.  The RG clarifies the ambiguities, which now largely lie in the choice of the field parameterizations
in the UV boundary conditions of the theory \cite{cth}.  

Note, moreover, that at this order the running of the Higgs mass does not involve the running
of the $\alpha_i$. The running of the $\alpha_i$ leads to the tricky problem of the regeneration
of the contact term, which we'll treat elsewhere \cite{HRelsewhere}. However, ultimately, a stable Higgs mass would likely 
necessitate an infrared fixed point of these RG equations.

\section{THE SCALE INVARIANT MIXED  HIGGS/$R^2$ MODEL}

We can readily construct a scale invariant version of the mixed Higgs/$R^2$ model \cite{Ferreira:2019zzx,Ghilencea:2019rqj} in which
the inflationary parameters are predicted to be close to 
the original Starobinsky model which is in good agreement with present 
observations . As before, 
to obtain the observed amplitude of the perturbations $f_0\sim 10^{-5}$ requires that the scaleron has mass of $O(10^{13})$ GeV.

The Scale Invariant action is :
\bea
S&=&\int\sqrt{-g}\Big(\frac{1}{2}g^{\mu\upsilon}\partial_{\mu}\phi
\partial_{\nu}\phi+\frac{1}{2}g^{\mu\upsilon}\partial_{\mu}H\partial_{\nu
}H-\frac{1}{12}\alpha_\phi\phi^{2}R-\frac{1}{12}\alpha_\eta\eta^{2}R-\frac
{1}{12}\alpha_HH^{2}R
\nonumber \\
&&
-\frac{\lambda}{4}\phi^{4}-\frac
{\xi}{4}\eta^{4}-\frac{\omega}{4}H^{4}\Big)
\label{scaleinvariantstar}
\eea
Here we again have the auxiliary field $\eta$, which can be integrated out to recover the
$R^{2}$ term as in the non-scale invariant model of Section II.

Note that, in the limit $\alpha_\phi=\alpha_H$ and $\lambda=\omega=0$,  the theory is $SO(2)$ invariant. In this limit the BEH scalar is massless and the symmetry is not broken by radiative corrections. As a result the
Higgs mass $\beta$-function must vanish in this limit. 

Note the action is classically invariant under the Weyl-scale transformation:
\bea
g_{\mu\nu }(x)&=&e^{2\epsilon}g_{\mu\nu }(x) \qquad g^{\mu\nu}(x)=e^{-2\epsilon}g^{\mu\nu}(x)
\nonumber \\
&&
\left(  \phi,\eta,H\right)  =e^{-\epsilon}\left(  \phi,\eta,H\right)  
\eea
This leads to a conserved Noether current:
\bea
K_{\mu}=\partial_{\mu}K \qquad
\makebox{where}
\qquad
K=\frac{1}{2}\Big(
(1-\alpha_\phi)\phi^{2}+(1-\alpha_H)H^{2}-\eta^{2}\Big) 
\eea
The current has the special form, in analogy to
topological currents, of being a derivative of a scalar,
$K$, which we refer to as the ``kernal.''  
A pre-Plankian phase will cause red-shifting and the conservation
of $K_{\mu}$ implies that  $K\rightarrow \overline{K}$ constant. This
``condensation'' of the kernal,  $K$, spontaneously breaks scale symmetry and
has been dubbed ``inertial symmetry breaking,'' \cite{Ferreira:2018itt},
since it makes no reference to a potential.
The phase transition is 
from a highly disordered state  to an ordered state
and occurs when
fluctuations in $K$ become small compared to its average VEV,
i.e.,  $\left|  \left\langle K^{2}\right\rangle -\overline{K}%
^{2}\right|  \ll\overline{K}^{2}.$

We can rewrite the original dynamical $K$ as
\bea 
2K=\left(  (1-\alpha_\phi)\widehat{\phi}^{2}+(1-\alpha_H%
)\widehat{H}^{2}-\widehat{\eta}^{2}\right)  e^{2\sigma/f}
\label{kernal}
\eea
where $f=\sqrt{2\overline{K}}$ canonically normalizes $\sigma$ in the broken phase,
and the dilaton, $\sigma$, apparently decouples in the Jordan frame \cite{Ferreira:2016kxi} but acquires derivative couplings to Standard  Model fields via contact terms or, formally, via the Weyl transformation going to the Einstein frame. The inertial symmetry
breaking can be viewed as the red shifting of the dilaton
to a constant VEV, $\sigma\rightarrow \sigma_0$,  and thus
$K\rightarrow \overline{K}$  (the VEV of the dilaton has no absolute meaning and is only defined
relative to the other field VEV's, and in fact the dilaton
itself is only well-defined in the ordered phase 
as a Nambu-Goldstone boson when $K\rightarrow \overline{K}$).

The resulting constraint is: 
\bea
2\overline{K} &=&(1-\alpha_\phi)\widehat{\phi}%
^{2}-\widehat{\eta}^{2}+(1-\alpha_H)\widehat{H}^{2}
= \makebox{constant.}
\label{Kernel}
\eea
In what follows we will determine the radiative corrections to the Higgs mass following from the action eq(\ref{scaleinvariantstar}). Dropping the dilaton, which does not contribute significantly to the radiative correction, the action becomes:
\bea
S&=&\int\sqrt{-g}\Big(\frac{1}{2}g^{\mu\upsilon}\partial_{\mu}\hat\phi
\partial_{\nu}\hat\phi+\frac{1}{2}g^{\mu\upsilon}\partial_{\mu}\hat H\partial_{\nu}\hat H
\nonumber \\
&&
+\frac{1}{2}M^{2}R\Big(-\frac{1}{6M^{2}}\alpha_\phi\hat\phi^{2}-\frac{1}{6M^{2}%
}\hat\eta^{2}-\frac{1}{6M^{2}}\alpha_H\hat H^{2}\Big)-\frac{\lambda}{4}\hat\phi
^{4}-\frac{\xi}{4}\hat\eta^{4}-\frac{\omega}{4}\hat H^{4}\Big)
\label{scaleSI}
\eea
In what follows, for notational convenience, all fields are understood to be hatted fields although we do not display the hats.
The constraint of eq(\ref{Kernel}) means that we can consider any
pair of the $\left(  \phi,\eta,H\right)  $ as unconstrained, while the
remaining field is then determined by the constraint to be a function of the 
other two.  The case that $\eta$ is eliminated manifestly preserves the underlying $SO(2)$ symmetry where  $\left(  \phi,H\right)  $ form a doublet. The case that $\phi$ is eliminated hides this symmetry but gives a calculation very similar to the non-scale invariant case. 
As a gratifying check on the analysis we have found that both routes lead to the same result for the BEH scalar mass.

\subsection{Calculation manifestly preserving the underlying $SO(2)$ symmetry structure}\label{SO(2)sect}

Defining $\overline{K}={3M^{2}}$  we presently
use the constraint, eq(\ref{Kernel}), to eliminate the $\eta^2$ term in the coefficient of the Ricci scalar in the action, eq(\ref{scaleSI}).  We thus obtain:
\be
S_A=\int\sqrt{-g}\left(\frac{1}{2}g^{\mu\upsilon}\partial_{\mu}\phi\partial_{\nu
}\phi+\frac{1}{2}g^{\mu\upsilon}\partial_{\mu}H\partial_{\nu}H+\frac{1}%
{2}M^{2}R\ \Omega^{2}-\frac{\lambda}{4}\phi^{4}-\frac{\xi}{4}\eta^{4}%
-\frac{\omega}{4}H^{4}\right)
\label{scaleStar}
\ee
where,
\bea
\Omega^{2}=\left(1-\frac{1}{6M^{2}}\phi^{2}-\frac{1}{6M^{2}}H^{2}\right).
\eea

Now, the contact interactions arising from graviton exchange must be incorporated into
the action \cite{Hill:2020oaj} and these again have the effect of forcing one into the Einstein frame. 
This means that the disorder-order phase
transition proceeds directly from pre-Plankian chaos to an Einstein frame
and there is no physical meaning to the ``Jordan frame.''
However, as in the NSI case, the form of the contact terms is
that of a Weyl transformation, even though the true metric remains  invariant. 
 
To incorporate the contact terms we therefore perform a formal Weyl transformation to the Einstein frame,
\bea
S=\int\sqrt{-g}\left(\frac{1}{2}\Omega^{-2}\partial_{\rho}\phi\partial^{\rho}%
\phi+\frac{1}{2}\Omega^{-2}\partial_{\rho}H\partial^{\rho}H+\frac{3M^{2}}%
{4}\Omega^{-4}\partial_{\rho}\Omega^{2}\partial^{\rho}\Omega^{2}+\frac{1}%
{2}M^{2}R\right.
\nonumber \\
\qquad\qquad\left.-\frac{\lambda}{4}\Omega^{-4}\phi^{4}-\frac{\xi}{4}%
\Omega^{-4}\eta^{4}-\frac{\omega}{4}\Omega^{-4}H^{4}\right)
\eea
We are interested in the one loop radiative 
analysis of the BEH mass which will parallel the analysis
of the scale non-invariant model.  The main effects
will come through the potential,  $\frac{\xi}{4}\Omega^{-4}\eta^{4}$
and we presently set $\lambda=\omega=0$.
The contribution of this term to the BEH mass is zero at the classical minimum of the
potential, but will be generated at one loop, order $\hbar.$

Following the Coleman-Weinberg procedure \cite{Coleman:1973jx} we shift the fields
to classical background VEVs, plus small quantum corrections
that are O$\left(  \sqrt{\hbar}\right)  $, and the leading one-loop result is
 then O$\left(  \hbar\right)  $.
In the following we define the parameters:
\bea
 \gamma_{i}=(1-\alpha_{i}), \qquad
\gamma=\frac{1}{2}\left(  \gamma_\phi+\gamma_{H}\right), 
\qquad \gamma^{\prime}=\frac{1}{2}\left(  \gamma_\phi-\gamma_{H}\right).  
\eea
From the constraint of eq.(\ref{Kernel}) we obtain:
\bea
V_0= \frac{\xi}{4}\Omega^{-4}\eta^{4}=9\xi M^4
\left(  1 -\gamma_\phi\frac{\phi^{2}}{6M^{2}}-\gamma_{H}\frac{H^{2}}{6M^{2}}\right)  ^{2}\left(1-\frac{\phi^{2}}{6M^{2}}-\frac{H^{2}}{6M^{2}}\right)^{-2}.
\label{pot}
\eea

As mentioned above, 
the kinetic terms of the scalars $\phi$ and $H$  are $SO(2)$
invariant, and so too the potential in the limit 
$\alpha_\phi=\alpha_{H}$, or $\gamma'=0$.
This will lead to an important constraint on
our results and simplify our calculation.
Since the kinetic terms are $SO(2)$ invariant we can
use a polar representation
of the fields, $\left(  \phi,H\right)  \rightarrow(\rho,\theta)$\ :
\bea
\phi^{2}+H^{2}=\rho^{2} \qquad \qquad \phi=\rho\cos\theta,\qquad H=\rho
\sin\theta .
\eea
As a notational convenience we set $6M^{2}=1$ and restore this factor at
the end of the calculation. Hence,
$
\Omega^{2}=(1-\phi^{2}-H^{2})\rightarrow
(1-\rho^{2})$, and
the classical field VEV's will be  $\rho_{0}$ and $\theta_{0}$,
\bea
\phi_{0}=\rho_{0}\cos\theta_{0} \qquad \makebox{ and}
\qquad H_{0}=\rho_{0}\sin\theta_{0}
\approx\rho_{0}\theta_{0}.
\eea
We will be interested in the BEH boson mass term, to order $H_{0}^{2}
\approx\left(  \rho_{0}\theta_{0}\right)  ^{2}$.
The ``small'' polar quantum fluctuations are ${\cal{O}}(\epsilon)=\sqrt{\hbar}$:
\bea
\rho=\rho_{0}+ \epsilon r;\qquad \theta=\theta_{0}+ \epsilon 
\vartheta.
\eea
The polar coordinates have the advantage of diagonalizing the kinetic terms.
Expanding the kinetic terms to O$\left(  \hbar\right)  $ we see:
\bea
S_{KT}&=&\int\sqrt{-g}\left(\frac{1}{2}\Omega^{-2}\partial_{\rho}%
\phi\partial^{\rho}\phi+\frac{1}{2}\Omega^{-2}\partial_{\rho}H\partial^{\rho
}H+\frac{3M^{2}}{4}\Omega^{-4}\partial_{\rho}\Omega^{2}\partial^{\rho}%
\Omega^{2})\right)
\nonumber \\
&\rightarrow &\int\sqrt{-g}\left(\frac{1}{2}\left(  \Omega_{0}^{-4}\right)
drdr+\frac{1}{2}\rho_{0}^{2}\Omega_{0}^{-2}\left(  d\vartheta d\vartheta
\right)  \right).
\eea
Then canonically normalizing the fields we have
\bea
S_{KT}
=\int\sqrt{-g}\left(\frac{1}{2}d\widehat{r}d\widehat{r}+\frac{1}%
{2}d\widehat{\vartheta}d\widehat{\vartheta}\right),
\eea
where,
\bea
&&  Z^{-1}r =  \widehat{r}, \qquad K^{-1}\vartheta=\widehat{\vartheta},
\qquad
Z = \Omega_{0}^{2}, \qquad K=\Omega_{0}\rho_{0}^{-1}, \qquad\Omega
_{0}^{2}=(1-\rho_{0}^{2}) .
\eea

Turning to the potential, we have:
\bea
\label{pot111}
V_0\rightarrow f\left(  \ 1-\gamma_\phi\left(  \rho\cos\theta\right)
^{2}-\gamma_{H}\left(  \rho\sin\theta\right)  ^{2}\right)  ^{2}(1-\rho
^{2})^{-2}
\eea
where $f={\xi}/{4}$.
We see that the $SO(2)$ symmetry is explicitly broken by non-zero $2\gamma'=\gamma_\phi-\gamma_H.$

We can go to the classical minimum of eq.(\ref{pot111}) and expand in
the normalized quantum fluctuations.
The classical minimum is a flat direction that corresponds to:
\bea
\label{Pmin11}
1-\rho_{0}^{2}\left(  \gamma_\phi+\left(  \gamma_{H}-\gamma_\phi\right)
\left(  \sin^{2}\theta_{0}\right)  \right)  =0
\label{min}
\eea
Since the potential is proportional to
the lhs squared, the classical minimum has
a vanishing energy and hence the Higgs mass term arises at $O(\hbar)$.

 The renormalized mass matrix squared in the  $\left(  \widehat{r}%
,\widehat{\vartheta}\right)  $ basis
is given by
\bea
\Xi^2=f\left(
\begin{array}
[c]{cc}%
Z^{2}\frac{d^{2}}{d\hat{r}^{2}}V_0 &  ZK\frac{d^{2}}{d\hat{r}d\hat\vartheta}V_0\\
ZK\frac{d^{2}}{d\hat{r}d\hat\vartheta}V_0 & K^{2}\frac{d^{2}}{d\hat\vartheta^{2}}V_0
\end{array}
\right)  
\eea
With the constraint of eq(\ref{min}) used to eliminate $\rho_0$, we find to $\ O\left(  \theta_0^{2}\right)  $:
\bea
\Xi^2=8f\left(
\begin{array}
[c]{cc}%
\gamma+\gamma^{\prime}-2\gamma^{\prime}\theta_0^{2} & -2\gamma^{\prime}\theta_0
X\\
-2\gamma^{\prime}\theta_0 X & 4\theta_0^{2}\frac{(\gamma^{\prime})^{2}}%
{\gamma+\gamma^{\prime}-1}%
\end{array}
\right)  +O\left(  \theta_0^{3}\right)  
\label{mass2}
\eea
where 
\bea
X=\left(  \allowbreak\frac{\gamma+\gamma^{\prime}}{\gamma
+\gamma^{\prime}-1}\right)  ^{1/2}
\eea
The trace of  $(\Xi)^4 $ is then obtained,
\bea
\makebox{Tr}(\Xi)^4 =\left(  64f^{2}\left(  \gamma+\gamma^{\prime}\right)  ^{2}-256f^{2}\left(
\gamma+\gamma^{\prime}\right)  \gamma^{\prime}\left(  \gamma-\gamma^{\prime
}-1\right)  \frac{\theta_0^{2}}{\gamma+\gamma^{\prime}-1}+\allowbreak O\left(
\theta_0^{4}\right)  \right) 
\eea
and the Higgs mass term is the $O\left(  \theta_0^{2}\right)  $ term.

We restore $6M^{2}$, and note
that the normalized Higgs field in the polar representation 
at the potential minimum,
eq.(\ref{Pmin11}), where $1-\gamma_\phi \rho_0\approx 0$,
is now:
\bea
h=\Omega_{0}^{-1}\rho_{0}\sin\theta_{0}\rightarrow K\theta_{0}%
=\frac{1}{\sqrt{\left(  \gamma+\gamma^{\prime}-1\right)
}}\theta_{0}
\eea
Using the Coleman Weinberg form for the induced potential, eq(\ref{EP}), the BEH mass term is given by:
\be
\Gamma
 = -\frac{3}{2\pi^{2}}M^{2}{\xi^{2}}%
\gamma^{\prime}\left(  \gamma+\gamma^{\prime}\right)  \left(  \gamma
-\gamma^{\prime}-1\right)  h^{2}\ln\left(  \Lambda^{2}/\mu^{2}\right)  
\ee
This vanishes in the $\gamma^{\prime}=0$ limit 
in which case the spontaneously broken $SO(2)$ symmetry
implies that the BEH scalar is a Nambu Goldstone boson
and must then have vanishing mass.
In terms of the original non-minimal couplings we have:
\bea
\Gamma=-\frac{3}{4\pi^{2}}{\xi}^{2}
\left(  1-\alpha_\phi\right)  \left(  \alpha_H-\alpha_\phi\right)
\alpha_HM^{2}h^{2}\ln\left(  \Lambda^{2}/\mu^{2}\right)  
\label{res1}
\eea

\subsection{Calculation of the Radiative Correction with Source Terms}

In the calculation of the previous section we worked with the action of eq(\ref{scaleStar}) evaluated at the classical potential minimum, eq(\ref{min}). As a result it was not necessary to include source terms and the renormalised mass matrix, eq(\ref{mass2}),  involved only a a single field that we took to be $\theta$.  More generally we may add sources for $H$ and  $\phi$ to the action which, as discussed in Section \ref{NSI},   allows us to study the effective potential as a function of both the classical fields $h$ and $\kappa$  where  $ H=h+\epsilon x,\;\;\;\chi=\kappa+\epsilon y$, unconstrained by the classical minimum condition eq(\ref{min}). 

The potential still has the form given in eq(\ref{pot}) and the unrenormalized mass squared mass matrix for the quantum fluctuations
$\ \left(  x,y\right)  $ is given by:
\be
\frac{1}{2}\left( {\begin{array}{*{20}{c}}
{y}&{ x}
\end{array}} \right)\Xi_{0}^{2}\left(  \
\begin{array}
[c]{c}%
y\\
x
\end{array}
\right),\qquad  
\Xi _0^2 = {\left. {\left( {\begin{array}{*{20}{c}}
{\frac{{{\partial ^2}V_0}}{{\partial {\phi ^2}}}}&{\frac{{{\partial ^2}V_0}}{{\partial \phi \partial H}}}\\
{\frac{{{\partial ^2}V_0}}{{\partial \phi \partial H}}}&{\frac{{{\partial ^2}V_0}}{{\partial {H^2}}}}
\end{array}} \right)} \right|_{\phi  = \kappa ,H = h}}.
\ee
Without the constraint of eq.(\ref{min})  the resulting form is algebraically lengthy and requires the use of Mathematica or Maple to evaluate it, so we do not quote the result here.

Turning to the normalisation of the fields, using:
\be
\frac{3M^{2}}{4}\Omega^{-4}\partial_{\rho}\Omega^{2}\partial^{\rho}\Omega^{2}=\Omega^{-4} \Big(\frac{1}{6}\phi^2(\partial\phi)^2+\frac{1}{6}H^2(\partial H)^2\Big)
\ee
the kinetic energy term may be rewritten as:
\bea
KE=&=&\frac{3M^{2}}{4}\Omega^{-4}\partial_{\rho}\Omega^{2}\partial^{\rho}\Omega^{2}+\frac{1}{2} \Omega^{2}\left(\partial^\mu\phi\partial_\mu\phi +\partial_\mu H\partial^\mu H\right)\nonumber\\
&=&\frac{1}{2}\Omega^{-4}\left((1-\frac{1}{6}H^2)(\partial\phi)^2+
(1-\frac{1}{6}\phi^2)(\partial H)^2+\frac{2}{6}\phi H\partial\phi\partial H\right).
\eea
Thus the kinetic energy has the form:
\be
\frac{1}{2}\left( {\begin{array}{*{20}{c}}
{\partial y}&{\partial x}
\end{array}} \right)Z\left( {\begin{array}{*{20}{c}}
{\partial y}\\
{\partial x}
\end{array}} \right),\;\;\;\;
\qquad
Z = \Omega {(\kappa,h )^{ - 4}}\left( {\begin{array}{*{20}{c}}
{1 - \frac{{{h^2}}}{6}}&{\frac{{h\kappa }}{6}}\\
{\frac{{h\kappa }}{6}}&{1 - \frac{{{\kappa ^2}}}{6}}
\end{array}} \right).
\ee
The renormalized physical mass$^{2}$ matrix is now:
\bea
\Xi_1^{2}=Z^{-1/2}\Xi_{0}^{2}Z^{-1/2},
\eea
which, inserted in eq(\ref{EP}), gives the one loop contribution to the quantum effective potential.

To compare with the previous result we can use this to calculate the BEH mass about the minimum of the potential which is given by $h=0$ and
${\phi}=\phi_0+\phi_1$, where $\phi_0=\sqrt{6/\alpha_\phi}$ is the minimum of $V_0$, and $\phi_1$ is a correction to the minimum due to the one loop correction, $V_1$, given by:
 \be
\phi_1= - \left.  \frac{\partial {V_1}}{\partial \phi }\left(
\frac{\partial ^2 V_0}{\partial {\phi ^2}} \right)^{-1} \right|_{\phi  =\phi_0 }
 \ee
At the minimum the radiative correction to the BEH mass at one loop order is  given by:
\bea
\delta m_{h^2}=-\frac{1}{768\pi^{2}}{f_0^4M^2}\left(  1-\alpha_\phi\right)  \left(  \alpha_H-\alpha_\phi\right)
\alpha_H h^{2}\ln\left(  \Lambda^{2}/\mu^{2}\right)  
\eea
agreeing with the result of eq.(\ref{res1}), with ${\ f_{0}^{2}}={24\xi}$, from 
$G=({3}/{8})M^{4}f_{0}^{2}=9M^{4}\xi$.

We have also done a calculation
in which we choose to eliminate $\phi$, rather that $\eta$,
using the constraint of eq(\ref{Kernel}). This has the advantage of rendering the calculation similar to that of the NSI case of Section II. While this method does spoil the {\em manifest} $SO(2)$ symmetry,
it yields the same result as  eq.(\ref{res1}) which
displays the $SO(2)$ symmetry as $\gamma'\rightarrow 0$.
The calculation also checks whether such a change of variables affects the result --- indeed we find it does not.
The details are rather lengthy and we won't present them here.   


\section{Conclusions}

We have provided detailed calculations of the low energy effective
theory, in particular, the BEH (``Higgs'') boson mass,
emerging from  mixed Higgs/$R^2$ inflation models.  We have considered
the standard non-scale-invariant form, and we have also introduced
a scale invariant form where the scale symmetry is broken
spontaneously inertially. 

A key point here is that the contact terms show that the theory formulated in the Jordan frame involving anomalous couplings of the scalars to the Ricci scalar is identical to the Einstein frame with only an Einstein Hilbert term, $M_P^2R$, and higher dimension operators.
The only ambiguity is the choice of source terms,
equivalently, the order parameters of the effective potential. In the original 
Higgs/$R^2$ model we argued the natural choice is $\chi$ and $H$, where $\chi$ is the field
with a natural canonical kinetic term. In the scale invariant case the choice is any two of three of the 
original set of fields, $\phi$, $\eta$ and $H$, introduced when defining the theory. These are constrained by the kernel of the Weyl current, $K$, which  develops a VEV that spontaneously breaks Weyl-scale invariance.

 Once the source terms are specified, the radiative corrections are unambiguous.
 Our results show that in both case the radiative corrections to the BEH mass do not
 vanish in the ``conformally coupled'' limit, $\alpha_H=1$, in contrast to the results
obtained by previous authors \cite{Salvio:2017qkx}.  

A renormalization group approach, introduced here, 
significantly improves our understanding of these theories
and expedites these analyses.  We have
only touched upon the RG approach here in  application to the
non-scale invariant Higgs/$R^2$ model.  In the RG approach
the choice of variables is embedded in the choice
of boundary conditions on the RG equations.
This will be developed elsewhere in greater detail \cite{HRelsewhere}.
Note that the running of the Higgs
boson mass is not due to the running of the $\alpha_i$
but rather mainly due to the effects of other relevant operators
that emerge from the non-polynomial Starobinsky
potential at the Planck-scale. At the single-log order discussed here,
the $\alpha_H$ enters in the UV only as a boundary condition
upon the relevant couplings. 

The determination of the radiative corrections to the BEH mass is of importance 
to the viability of the Starobinsky inflationary model. To generate an acceptable 
period of inflation  the model requires a very heavy scaleron, of $O(10^{13})GeV,$ and
the coupling of the scaleron to the BEH scalar typically gives an unacceptably large
contribution to its mass leading to a severe hierarchy problem.  Conceiveably,
the $SO(2)$ symmetry of the scale invariant model, or its generalization, might allow 
the possibility of a protected BEH scalar mass, as a pseudo-Nambu-Goldstone boson.
We will discuss the implications of our results 
for the hierarchy problem in
detail elsewhere \cite{FHNR}.

{ NOTE ADDED:  After completion of this work we became aware of
the interesting papers of C. Steinwachs, et.al. which 
address the loop level consistency issues in these
models \cite{Steinwachs}.  We believe that 
the gravitational contact terms resolve many of
the issues  encountered by these authors. 
Moreover, the Wilson-style renormalization group approach exhibited 
in section II.D  provides
a simpler calculational framework and
will be developed further elsewhere.
}

\newpage


\appendix

\section{Regularized Loop Integrals and Quantum Scale Breaking}\label{loops}

We are interested in loop induced effective potentials.
These are contained in the log of the path integral:  $\ \Gamma=i\ln P$.
In the case of  a real scalar field of physical mass, $m$,
we have:
\bea
P=\underset{k}{\prod}\left(  k^{2}-m^{2}\right)^{-1/2}  
=\det\left(  k^{2}-m^{2}\right)^{-1/2}
\eea
hence,
\bea
\label{A2}
\Gamma=i\ln P= -\frac{i}{2}  \int\frac{d^{4}k}{\left(  2\pi\right)  ^{4}}%
\ln\left(  k^{2}-m^{2}+i\epsilon\right)  
\eea
This can be evaluated with a Wick rotation and Euclidean
momentum space cut off:
\bea
\label{A3}
\Gamma &= & \frac{1}{2}\int_0^{\Lambda}\frac{d^{4}k_E}{\left(  2\pi\right)^{4}}
\ln\left(  \frac{k_E^{2}+m^{2}}{\Lambda^{2}}\right) + (\makebox{irrelevant constants})
\nonumber \\
&=&
\frac{1}{64\pi^{2}}\left(  \ln\frac{\Lambda^{2}+m^{2}}%
{\Lambda^{2}}\Lambda^{4}-m^{4}\ln\frac{\Lambda^{2}+m^{2}}{\Lambda^{2}}
-\frac{1}{2}\allowbreak\Lambda^{4}+\Lambda^{2}m^{2}+m^{4}\ln\frac{m^{2}
}{\Lambda^{2}}\right)  
\eea
The
cutoff can be viewed is a spurious parameter, introduced to make the integral finite
and not part of the defining
action. 
The only physically meaningful dependence upon $\Lambda$ is contained 
in the
logarithm, where it reflects scale symmetry breaking by the 
quantum trace anomaly. Powers of $\Lambda$, e.g., $\Lambda^{4},\Lambda^{2}m^{2}$. spuriously break scale
symmetry and are not part of the classical action  \cite{Bardeen}.

It is therefore conceptually useful to have a definition of the
loops in which the spurious powers of $\Lambda$ do not arise.
This can be done by defining the loops
applying projection
operators on the integrals.  The projection operator
\bea
P_{n}=\left(  1-\frac{\Lambda}{n}\frac{\partial}{\partial\Lambda}\right)  
\eea
removes any terms proportional to $\Lambda^{n}.$ 
Since the defining classical Lagrangian
has mass dimension 4 and involves no terms with  $\Lambda^{2}m^{2}$ or
$\Lambda^{4},$ we define the regularized loop integrals as:
\bea
\label{A5}
\Gamma &\rightarrow &\frac{1}{2}P_{2}P_{4}\int_0^{\Lambda}\frac{d^{4}k_E}{\left(  2\pi\right)^{4}}
\ln\left(  \frac{k_E^{2}+m^{2}}{\Lambda^{2}}\right) + (\makebox{irrelevant constants})
\nonumber \\
&=&
-\frac{1}{64\pi^{2}} \left(m^{4}\left(  \ln\frac{\Lambda^{2}}{m^{2}%
}-1\right)  +O\left(  \frac{m^{6}}{\Lambda^{2}}\right)  \right) 
\eea
where we take the limit $\Lambda>>m$ to suppress $O\left( m^{6}/
{\Lambda^{2}}\right)  $ terms.
It appears this can be consistently used in $N$-loop
calculations with $\left(  P_{2}P_{4}\right)  ^{N}$ though 
we now apply it only to single loop amplitudes. 

Eq.(\ref{A5})  defines the quantum effective potential
for a classical real scalar field with mass term
$V_c=m^2\phi^2/2$:
\bea
\Gamma &=& V_c
-\frac{1}{64\pi^{2}}m^{4} \ln\frac{M^{2}}{m^{2}%
} 
\eea
where we have replaced the cut-off by the Planck-scale.

As a check we can compute the $\beta$-function for a quartic
coupling constant. Consider $V_c= \lambda\phi^4/4$. We expand $\phi$
in a classical plus quantum fluctuation field $x$,  $\phi\rightarrow \phi_c+ \epsilon x$
and $V_c\rightarrow  \lambda\phi_c^4/
4 + 3\hbar \lambda\phi_c^2x^2/2$, where terms linear in $x$ integrate to zero.  We see that the physical mass of $x$ is $m^2= 3\lambda\phi_c^2$
and using $\Gamma$ we have:
\bea
\Gamma &=& \frac{\lambda}{4}\phi_c^4
-\frac{9\lambda^2}{64\pi^{2}} \phi_c^4 \ln\frac{M^{2}}{m^{2}%
} 
\eea
Therefore, we see that $\lambda$ runs as:
\bea
\lambda(\mu)=\lambda_0
- \frac{9\lambda^2 }{8\pi^{2}} \phi^4 \ln\frac{M}{\mu
} 
\eea
where the initial value at the Planck-scale is $\lambda_0$
and the $\beta$-function is:
\bea
\beta = \frac{\partial\lambda(\mu)}{\partial \ln(\mu)} =\frac{9\lambda^2}{8\pi^{2}}
\eea
a well-known result (see ref.\cite{Coleman:1973jx,cth} and references therein). 

The above result is ${\cal{O}}(\hbar)$.
since the expansion in $\hbar$ is an expansion in the number of loops.
Note that $\Gamma =i\hbar  \ln(i S/\hbar)$, so a classical action
$S$ produces an O($1$) result in the $\hbar$ expansion.  The
quantum field
kinetic terms are $\sim  S\sim \int\hbar (\partial \phi)^2$ hence
a propagator is $1/\hbar$ and a 
 single Feynman loop is $\Gamma \propto \hbar  \ln(i \int \hbar(\partial \phi)^2/\hbar)
 \sim \hbar$; $N$ loops are $\propto \hbar^N$.

\noindent
\section{Equivalence of On-shell Field Configurations and 
Ambiguities in Effective Potentials}\label{onshell}

It is fairly easy to give a formal proof of the equivalence of different field
choices for calculation of effective potentials
with on-shell classical background fields. Our proof 
for $N=1$ fields is schematic 
and readily can be generalized to $N$ fields.
Consider the action: 
\bea 
S=\int\frac{1}{2}\partial\phi\partial\phi-V\left(  \phi\right)  
\eea
For an effective potential, such as a Coleman-Weinberg calculation,
 $V(\phi)  $ is understood to contain the source
terms.  
Note that we expand in a classical background
field $\phi_0$ plus a quantum fluctuation, $x$,
$\phi=\phi_{0}+\epsilon x$
(where $\epsilon = \sqrt{\hbar})$, and we have,
\bea
\int\frac{1}{2}\partial\phi_{0}\partial\phi_{0}+\partial\phi_{0}\partial
x+\frac{1}{2}\partial x\partial x-V\left(  \phi_{0}\right)  -\frac{d}%
{d\phi_{0}}V\left(  \phi_{0}\right)  x-\frac{1}{2}\frac{d^2}{d\phi_{0}^2}V\left(  \phi_{0}\right)  x^{2}
\eea
We now make the assumption that $\phi_{0}$ is ``on-shell,''
i.e., satisfies its classical equation of motion:
\bea
\partial^{2}\phi_{0}=-\frac{d}{d\phi_{0}}V\left(  \phi_{0}\right) 
\eea
hence, integrating by parts,
\bea
S= \int\frac{1}{2}\partial\phi_{0}\partial\phi_{0}+\frac{1}{2}\partial x\partial x-V\left(  \phi_{0}\right)  -\frac{1}{2}(\frac{d^2}{d\phi_{0}^2}V\left(  \phi_{0}\right))  x^{2}
\eea
If we are only interested in  a potential, we 
simplify by assuming $\phi_{0}=$ spatially (and temporally) constant. 
Therefore the ``on-shell'' condition
requires: 
\bea
\frac{d}{d\phi_{0}}V\left(  \phi_{0}\right)  =0
\eea
i.e.,  $\phi_{0}$ must be at the minimum of the potential.
Bear in mind that the source currents shift the true potential minimum to
an arbitrary value, $\phi_0$, and the potential we compute will
correspond to the energy of a state with lowest energy subject
the constraint that the expectation value of $\phi$ is $\phi_0$.

Now suppose $\phi\left(  \chi\right)  $ is a function of a new
field $\chi.$ 
Then
\bea
S=\int\frac{1}{2}\partial\phi\partial\phi-V\left(  \phi\right)
\rightarrow\int\frac{1}{2}\left(  \frac{\partial\phi}{\partial\chi}\right)
^{2}\partial\chi\partial\chi-V\left(  \phi\left(  \chi\right)  \right)  
\eea
and now we want the expansion $\chi=\chi_{0}+\epsilon y$ with spatially
constant $\chi_{0}$
\bea
S&=& \int\frac{1}{2}\partial\phi\partial\phi-V\left(  \phi\right)
\rightarrow\int\frac{1}{2}\left(  \frac{\partial\phi_{0}}{\partial\chi_{0}%
}\right)  ^{2}\partial y\partial y-V\left(  \phi_{0}\right)  -\frac
{\partial\phi_{0}}{\partial\chi_{0}}\frac{d}{d\phi_{0}}V\left(  \phi
_{0}\right)  y
\nonumber \\  &&
\qquad\qquad-\frac{1}{2} \left(  \frac{\partial\phi_{0}}{\partial\chi_{0}%
}\frac{d}{d\phi_{0}}\left(  \frac{\partial\phi_{0}}{\partial\chi_{0}}\frac
{d}{d\phi_{0}}\right)  V\left(  \phi_{0}\right)  \right)  y^{2}
\eea
Examine the last term, which takes the form
\bea
-\frac{1}{2}\left(  \frac{\partial\phi_{0}}{\partial\chi_{0}}\right)
^{2}\left(  \frac{d^2}{d\phi_{0}^2}\left(  V\left(  \phi
_{0}\right)  \right)  \right)  x^{2}
-\frac{1}{2}
\frac{\partial^{2}\phi_{0}}{\partial\chi_{0}^{2}}\left(\frac{\partial\phi_{0}}{\partial\chi_{0}}
\frac{\partial}{\partial\phi_{0}} V\left(  \phi_{0}\right)  \right)
x^{2}
\eea
The last term is problematic off-shell, but on-shell we have:
\bea
\frac{\partial\phi_{0}}{\partial\chi_{0}}\frac{\partial}{\partial\phi_{0}%
}\left(  V\left(  \phi_{0}\left(  \chi_{0}\right)  \right)  \right)
=\frac{\partial}{\partial\chi_{0}}\widehat{V}\left(  \chi_{0}\right)  =0
\eea
That is, $\chi_{0}$  must be a minimum of the new potential: 
\bea
\widehat
{V}\left(  \chi_{0}\right)  =V\left(  \phi_{0}\left(  \chi_{0}\right)
\right)  
\eea
which corresponds to the minimum of $V\left(  \phi_{0}\right)  $.
Hence the action is:
\bea
S\rightarrow\int\left(  \frac{1}{2}\left(  \frac{\partial\phi_{0}}%
{\partial\chi_{0}}\right)^{2}\partial y\partial y-\widehat{V}\left(
\chi_{0}\right)  -\frac{1}{2}\left(  \frac{\partial\phi_{0}}%
{\partial\chi_{0}}\right)^{2}  \frac{d^2}{d\phi^2_{0}%
}  {V}\left(  \phi_{0}   \right)  y^{2}\right)  
\eea
Renormalizing $y$ yields:
\bea
\int\left(  \frac{1}{2}\partial y\partial y-\widehat{V}\left(  \chi
_{0}\right)  -\frac{1}{2}
  \frac{d^2}{d\phi_{0}^2}
{V}\left(  \phi_{0}  \right)  y^{2}  \right)
\eea
Since
\bea
\frac{1}{2}\frac{d^2}{d\phi_{0}^2}
V\left(  \phi_{0}\right)   =m^{2}
\eea
is common to both the $\chi$ and $\phi$
theories, therefore the quantum
potentials, $\propto \int \ln(k^2 +m^2)$
must be equivalent when the background
fields are localized at the minimum of the potential (including
source terms).

We can test this theorem in a simplified model (we will use the model in Appendix \ref{ambiguity}
to illustrate the ambiguities that result if there are different Weyl frames).
We consider a model which is similar
to those encountered in Starobinsky inflation.
It has the fields $\Omega,H$ and external sources $\left(
J,K\right) $ with action:
\bea
S=\int\; \frac{1}{2}\partial H\partial H+\frac{1}{2}M^{2}\partial
\ln\Omega\partial\ln\Omega\ \ -V(H,\Omega)-JM\ln\left(  \Omega\right) -KH
\eea
where 
\bea
\label{pott}
V(H,\Omega)=\frac{1}{4}M^{4}\left(  \Omega^{2}-\left(  1+\frac
{\alpha}{M^{2}}H^{2}\right)  \right)  ^{2}
\eea
We use the representation of the field, $\Omega\left(  \chi\right) 
=\exp(\chi/M)$, and define classical background fields and
quantum fluctuations $\sqrt{\hbar}=\epsilon$,
$H=h+\epsilon x$, and $\chi=\kappa+\epsilon y$.

Our procedure is as follows:

\noindent
 (1) Expand the action to
O$\left(  \hbar\right)  $ including sources,
\bea
V(H,\Omega)-JM\ln\left(  \Omega\right)  -KH\rightarrow
W(J,K;h,\kappa,\epsilon x,\epsilon y)
\eea
(2) Determine on-shell conditions for $\left(  J,K\right)  $ from the
$O(\epsilon^0)$ (classical
term)
of the potential by imposing the
minimum conditions on $\left(  h,\kappa\right)  .$
\bea
\frac{d}{d\kappa}W(J,K;h,\kappa,0,0)=0\qquad\qquad
\frac{d}{dh}W(J,K;h,\kappa,0,0)=0
\eea
This determines $J,K$ as functions of $\left(  h,\kappa\right).$

\noindent
(3) Define the effective potential by performing the Legendre transformation
\bea
W(J,K;h,\kappa,\epsilon x,\epsilon y)-\frac{dW}{dJ}\kappa-\frac
{dW}{dK}h=\Gamma(\kappa,h,,\epsilon x,\epsilon y)
\eea
At this stage the terms linear in $\epsilon x,\epsilon y$ cancel. 
Hence the functions $J,K$ completely
drop out if the source terms are linear
in $x,y.$    Otherwise, $J,K$ will enter the quadratic and higher terms.

\noindent
(4) Expand $\Gamma(\kappa,h,,\epsilon x,\epsilon y)$ to determine the mass$^{2}$
matrix of the 
quantum fluctuations $\epsilon x,\epsilon y$ at order $\epsilon^{2}$
\bea
\Gamma(\kappa,h,,\epsilon x,\epsilon y)=\frac{1}{2}\left(
\begin{array}
[c]{c}%
y\\
x
\end{array}
\right)  ^{T}\Xi^{2}\left(  \kappa,h\right)  \left(  \
\begin{array}
[c]{c}%
y\\
x
\end{array}
\right) 
\eea
and the kinetic terms and renormalization constant matrix
\bea
\frac{1}{2}\left(
\begin{array}
[c]{c}%
\partial y\\
\partial x
\end{array}
\right)  ^{T}Z\left(  \kappa,h\right)  \left(  \
\begin{array}
[c]{c}%
\partial y\\
\partial x
\end{array}
\right)  
\eea
(5) 
Integrating out $(\epsilon x,\epsilon y)$
the quantum effective potential is now determined,
\bea
-\frac{1}{64\pi^{2}}\makebox{Tr}\left(  \left(  Z\Xi^{2}\left(  \kappa,h\right)
\right)  ^{2}\right)  \ln\left(  \Lambda^{2}/m^{2}\right)  
\eea
We apply this to two examples with different quantum field
 definitions.

\noindent
{\bf Example (1):}
With  $\Omega=\exp\left(\chi/M\right)$, the
potential $V( H, \Omega)$ in eq.(\ref{pott}),
and source terms  $-JM\ln
\left(  \Omega\right)  -KH$, and we 
follow the procedure, find that $J,K$ cancel in  
the mass matrix
\bea
\Xi^{2}&=&\frac{1}{2}\left(
\begin{array}
[c]{cc}%
6\kappa M-\alpha h^{2}+6\kappa^{2}+M^{2}\allowbreak & -\alpha hM-2\alpha
h\kappa\\
-\alpha hM-2\alpha h\kappa & -\alpha\kappa M+\frac{3}{2}\alpha^{2}h^{2}%
-\alpha\kappa^{2}%
\end{array}
\right)  \left(  \
\begin{array}
[c]{c}%
y\\
x
\end{array}
\right)  
\nonumber \\
Z^{-1}&=&\left(
\begin{array}
[c]{cc}%
1\allowbreak & 0\\
0 & 1
\end{array}
\right)  \left(  \
\begin{array}
[c]{c}%
y\\
x
\end{array}
\right)  
\eea
Keeping only quadratic order in $\kappa$, $h$ we find:
\bea
\makebox{Tr}\left(  Z^{-1}\Xi^{2}\right)  ^{2}=\left(  \frac{1}{4}M^{4}+3M^{3}%
\kappa\right)  +\left(  \frac{1}{2}\alpha^{2}M^{2}-\frac{1}{2}M^{2}%
\alpha\right)  \allowbreak h^{2}+O\left(  h^{4},\kappa^{2},\kappa
h^{2}\right)  
\eea

\noindent
{\bf Example (2)} Now consider the alternative parametrization $(\tilde \chi,H)$ where $\chi=M \ln(1+{\tilde\chi/K})$.  We keep the same sources
 $-JM\ln\left(
\Omega\right)$ and $V -KH$.  We now need to expand the log,
and we find that $J$
appears in quadratic terms of $(x,y)$  due
to nonlinear term  $J\tilde\chi^{2}$:
\bea
\Xi^{2}&=&\frac{1}{2}\left(
\begin{array}
[c]{cc}%
\allowbreak\allowbreak4\kappa M-\alpha h^{2}+4\kappa^{2}+M^{2}-\frac{1}%
{M}\kappa\alpha h^{2}-\frac{1}{M^{2}}\kappa^{2}\alpha h^{2} & -\alpha
h\kappa-\alpha hM\\
-\alpha h\kappa-\alpha hM & -\alpha\kappa M+\frac{3}{2}\alpha^{2}h^{2}%
-\frac{1}{2}\alpha\kappa^{2}%
\end{array}
\right)  \left(  \
\begin{array}
[c]{c}%
y\\
x
\end{array}
\right)  
\nonumber \\
Z^{-1}&=&\left(
\begin{array}
[c]{cc}%
\left(  1+\frac{2}{M}\kappa\right)   & 0\\
0 & 1
\end{array}
\right)  
\eea
we obtain
$\makebox{Tr}\left(  Z^{-1}\Xi^{2}\right)  ^{2}=\left(  \frac{1}{4}M^{4}+3M^{3}%
\kappa\right)  +\left(  \frac{1}{2}\alpha^{2}M^{2}-\frac{1}{2}M^{2}%
\alpha\right)  \allowbreak h^{2}+O\left(  h^{4},\kappa^{2},\kappa
h^{2}\right)  $
These expressions agree and yield the same potential
though the intermediate steps are quite different.
The equivalence is a consequence of the above 
theorem.

\section{Source Ambiguities}\label{ambiguity}

The equivalence proved in  Appendix \ref{onshell} requires that the same sources should be used in both parameterisations. Thus in Example (2) above it was important to choose the same source, $-JM \ln(\Omega)\equiv -J\chi$, an in Example (1),
and the result is independent of the integration over quantum fluctuations.
However, if instead we used the source $-J\tilde\chi$  we have changed
the order parameter and the two calculations will differ. Such an ambiguity is inherent to the theory and requires the choice of ``reasonable" sources when defining the theory. 

To illustrate this we note that in many studies of Higgs/$R^2$ inflation the anomalous coupling to the BEH field is not included when Weyl transforming to a pseudo Einstein frame. 
This residual anomalous coupling can be eliminated by a second Weyl transformation that completes the definition of the model in the Einstein frame.  Weyl transformations satisfy the group property and the two Weyl transformations are equivalent to the single Weyl transformation going directly to the Einstein frame.
 However the fundamental variables suggested by this approach differ from that of the previous section. 
 
The two stage model starts with rewriting the action of eq(\ref{action1})  in the form \footnote{Here, to keep the algebraic complexity to a minimum while demonstrating the ambiguity, we do include a component of the BEH scalar in the first Weyl transformation. This component vanishes in the conformal limit.}
\be
\label{R5S}
   S = \int d^4x \sqrt{- \tilde   g} 
        \left(  \frac{M}{2}\Omega_1^{-2}\;R(    g)
       -\frac{\xi}{4}\eta^4+\frac{1}{2}\partial_\mu     H\partial^\mu     H-\frac{1}{12}    H^2 R(    g)-V(    H)  \right)
   \ee
   where,
   \bea
   \Omega_1^{-2}
   = 1-\frac{\alpha_\eta}{6}\left(\frac{\eta}{M}\right)^2-\frac{\alpha_H-1}{ 6}\left( \frac{H}{M}\right)^2\equiv 
   \exp \Big(\sqrt{2\over 3}\frac{\chi}{ M}\Big).
   \label{omega1}
   \eea
   We perform the first Weyl transformation:
 \be
   \tilde g= \Omega_1^2  g_1,
\ee
 giving,
\be
 S=\int d^4x \sqrt{- g_1} 
        \left(\frac{M}{2} R(g_1)+\frac{1}{2}\partial_\mu \chi\partial^\mu \chi+\frac{1}{2}\partial_\mu \tilde H\partial^\mu \tilde  H-\frac{1}{12} \tilde H^2 R( g_1)-\frac{\xi}{4}\Omega_1^4 \eta^4-\Omega_1^4V( H)\right),  
 \ee
  in terms of the conformally rescaled field,
   \be
 H_1=\Omega_1H.
   \ee
   The second Weyl transformation eliminates the residual anomalous coupling of the Higgs field and is given by, 
   \be
   g_1=\Omega_2^2 g_2,
   \ee
   with the conformal factor,
   \be
   \Omega_2^{-2}=1-\frac{1}{6}\left({ H_1\over M}\right)^2\equiv
   \exp\Big({\sqrt{{2\over 3}}{\rho\over M}}\Big),
   \label{omega2}
   \ee
  giving:
\bea
 S&=&\int d^4x \sqrt{- g_2} 
\Big(\frac{M}{2} R(g_2)+\frac{1}{2}\Omega_2^2\partial_\mu \chi\partial^\mu \chi+\frac{1}{2}\partial_\mu \rho\partial^\mu \rho+\frac{1}{2}\Omega_2^2\partial_\mu H_1\partial^\mu  H_1
\nonumber \\
&&
\qquad\qquad
        -\frac{\xi}{4}\Omega_2^4\Omega_1^4 \eta^4-\Omega_2^4\Omega_1^4V( H)\Big)  
        \nonumber\\
        &=&\int d^4x \sqrt{- g} 
        \left(\frac{M}{2} R(g)+\frac{1}{2}\Omega_2^2\partial_\mu \chi\partial^\mu \chi+\frac{1}{2}\Omega_2^4\;\partial_\mu  H_1\partial^\mu   H_1-\frac{\xi}{4}\Omega^4\; \eta^4-\Omega^4 V(H)\right)  .
 \eea
 Since the Weyl transformations form a group, comparing to the case studied in Section \ref{NSI1} we have $g_2=g$ and $\Omega=\Omega_2\Omega_1$.
 Comparing with eq(\ref{action1}) we see that the difference is that now the natural choice of fundamental variables is  $\chi$ and $\tilde H$.  Proceeding as in Section II we obtain,
 \bea
\Gamma =\frac{1}{768\pi^{2}}M^{2}f_{0}^{4}\Big((\alpha_H-1)(\alpha_H-2)+1\Big)  h^{2}\left(  \ln\left(  \Lambda^{2}/m^{2}\right)
\right)  
\eea
A comparison with eq(\ref{mt1}) shows that the mass terms differ, demonstrating the ambiguity associated with the choice of fundamental fields. 

The resolution of this ambiguity follows from the existence of the contact terms. As stressed in \cite{Hill:2020oaj}, when the BEH scalar is included, this approach implicitly requires the inclusion of contact terms to take account of its anomalous coupling to the Ricci scalar. Including this term automatically takes one to the Einstein frame
so there is no meaning to the two stage Weyl transformations just discussed and there  is no corresponding ambiguity.

\newpage
\vskip 0.5 in
\noindent
 {\bf Acknowledgements}
\vspace{0.1in}

We thank P. Ferreira and J. Noller for discussions.  
This manuscript has been authored in part by Fermi Research Alliance, LLC under Contract No. DE-AC02-07CH11359 with the U.S. Department of Energy, Office of Science, Office of High Energy Physics.

\end{document}